\documentclass[lettersize,journal]{IEEEtran}

\usepackage{amsmath,amssymb,amsfonts}
\usepackage{amsthm}
\usepackage{graphicx}
\usepackage{cite}
\usepackage{textcomp}
\usepackage{array}
\usepackage{tabularx}
\usepackage{multirow}
\usepackage{makecell}
\usepackage{xcolor}

\usepackage[caption=false,font=normalsize,
labelfont=sf,textfont=sf]{subfig}

\usepackage{geometry}
\usepackage{algorithm}
\usepackage{algpseudocode}

\usepackage{stfloats}

\begin{document}

	\title{Secrecy Sum-Rate Maximization for RSMA-Enabled Pinching-Antenna Systems}

	\author{IEEE Publication Technology,~\IEEEmembership{Staff,~IEEE,}
	}

\author{Code Kebe, Saeid Pakravan, Wessam Ajib

\thanks{C. Kebe, S. Pakravan, and W. Ajib are with the Department of Computer Science, University of Quebec in Montreal (UQAM), Montreal, QC, Canada (e-mail: kebe.code@courrier.uqam.ca; pakravan.saeid@uqam.ca; ajib.wessam@uqam.ca).}

}

\maketitle

\begin{abstract}
This paper investigates secrecy sum-rate (SSR) maximization in a
downlink rate-splitting multiple access (RSMA)-enabled single-waveguide
pinching-antenna system (PASS) with an external eavesdropper. The RSMA
power allocation, secure common-rate allocation, and pinching-antenna
(PA) positions are jointly optimized subject to transmit-power,
per-user secrecy-rate, and PA-placement constraints. To tackle the
resulting nonconvex problem, an alternating-optimization framework is
developed, combining successive convex approximation for RSMA resource
allocation with feasibility-preserving projected gradient ascent for
PA-position optimization. Numerical results show that the proposed
design consistently outperforms the considered benchmarks in terms of
average SSR, demonstrating the benefits of joint RSMA resource and
PA-position optimization for secure PASS transmission.
\end{abstract}

\begin{IEEEkeywords}
Pinching-antenna systems (PASS), rate-splitting multiple access (RSMA), physical-layer security, secrecy sum rate.
\end{IEEEkeywords}

\section{Introduction}

Pinching-antenna systems (PASS) have recently emerged as a
reconfigurable wireless architecture in which radiating elements are
deployed along dielectric waveguides to shape the effective wireless
channels \cite{11222687}. Owing to the combined guided-wave and
free-space propagation, the pinching-antenna (PA) locations provide an
additional degree of freedom for channel reconfiguration. This feature
has stimulated growing interest in PA configuration and resource
allocation for multiuser transmission
\cite{11625963, ref_conference,11029492,11568577}.

The reconfigurability of PASS is particularly attractive for secure
multiuser transmission, as PA positioning can favorably shape the
legitimate channels relative to that of an eavesdropper
\cite{11202497,11205176}. Meanwhile, RSMA provides flexible
inter-user interference management through common--private signaling
and has been investigated for secure transmission in reconfigurable
wireless environments \cite{9963667}. More recently, RSMA has been
integrated with PASS for non-security applications
\cite{11558589,11597999,Pakravan2026PASSRSMA}.  This naturally raises
the question of how RSMA and PA reconfiguration should be jointly
designed when both multiuser interference and information leakage must
be controlled.

Addressing this question is nontrivial in a single-waveguide,
single-feed PASS. In the considered architecture, the PAs do not have
independent digital beamforming weights; they radiate the same
composite RSMA waveform, while spatial selectivity is created through
their positions. Thus, the RSMA resource allocation governs the
common--private transmission, whereas the PA positions simultaneously
shape the legitimate and eavesdropping channels. Their joint design
therefore directly governs the achievable secrecy rates.
Table~\ref{tab:comparison} positions the considered problem with
respect to closely related works.

\begin{table}[t]
\caption{Comparison With Closely Related Works}
\label{tab:comparison}
\centering
\scriptsize
\setlength{\tabcolsep}{2.4pt}
\renewcommand{\arraystretch}{1.1}
\begin{tabularx}{\columnwidth}{
@{}
>{\centering\arraybackslash}p{0.085\columnwidth}
>{\centering\arraybackslash}p{0.090\columnwidth}
>{\centering\arraybackslash}p{0.105\columnwidth}
>{\centering\arraybackslash}p{0.07\columnwidth}
>{\centering\arraybackslash}p{0.160\columnwidth}
>{\centering\arraybackslash}p{0.105\columnwidth}
>{\centering\arraybackslash}X
@{}}
\hline
\textbf{Ref.} &
\textbf{PASS} &
\textbf{RSMA} &
\textbf{PLS} &
\textbf{Power Alloc.} &
\textbf{PA Pos.} &
\textbf{Objective} \\
\hline

\cite{11029492}
& \checkmark & $\times$ & $\times$ & \checkmark & \checkmark
& Sum rate \\

\cite{11202497}
& \checkmark & $\times$ & \checkmark & \checkmark & \checkmark
& Secrecy rate \\

\cite{11205176}
& \checkmark & $\times$ & \checkmark & \checkmark & \checkmark
& Secrecy enhancement \\

\cite{9963667}
& $\times$ & \checkmark & \checkmark & \checkmark & $\times$
& Max--min secrecy rate \\

\cite{11558589}
& \checkmark & \checkmark & $\times$ & \checkmark & \checkmark
& Max--min rate \\

\cite{11597999}
& \checkmark & \checkmark & $\times$ & \checkmark & \checkmark
& Sensing/QoS \\

\cite{Pakravan2026PASSRSMA}
& \checkmark & \checkmark & $\times$ & \checkmark & $\times$
& Max--min rate \\
\hline

\textbf{Ours}
& \checkmark & \checkmark & \checkmark & \checkmark & \checkmark
& \textbf{SSR} \\
\hline
\end{tabularx}

\vspace{0.3ex}
\parbox{\columnwidth}{\scriptsize
\textbf{PLS}: Physical- layer security;
\textbf{Power Alloc.}: power allocation;
\textbf{PA Pos.}: PA positioning.}
\end{table}

Motivated by the above, we investigate secrecy sum-rate (SSR)
maximization for a downlink RSMA-enabled single-waveguide PASS in the
presence of an external eavesdropper. We jointly optimize the RSMA
power allocation, secure common-rate allocation, and PA positions
subject to transmit-power, per-user secrecy-rate, and PA-placement
constraints. The common stream is kept undecodable
at Eve to interfere with private-stream interception. To solve the resulting nonconvex problem,
we develop an alternating-optimization (AO) framework combining successive
convex approximation (SCA) for RSMA resource allocation with
feasibility-preserving projected gradient ascent (PGA) for PA-position
optimization. Numerical results demonstrate consistent SSR gains over
the considered benchmarks.

\section{System Model and Problem Formulation}
\label{sec:system_model}

\begin{figure}[]
    \centering
    \includegraphics[width=7.5cm,height=3.5cm]{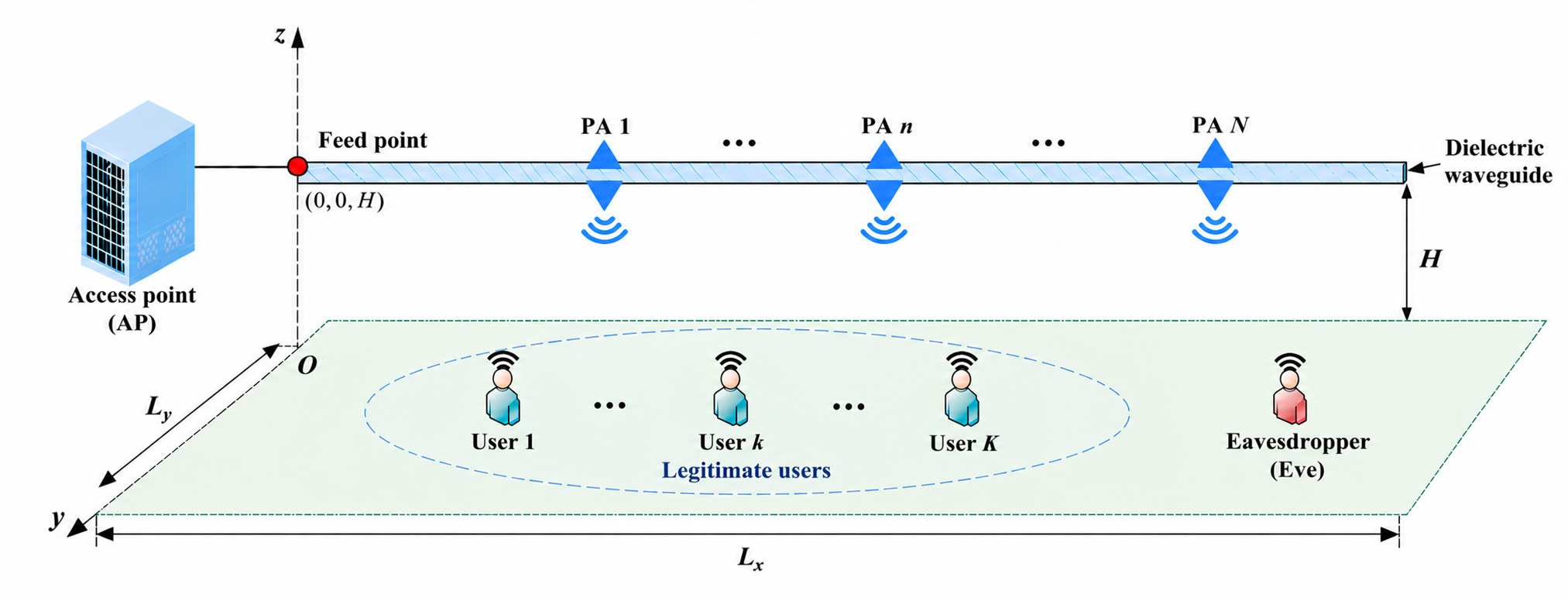}
 \caption{Secure RSMA-enabled single-waveguide PASS system.}
    \label{fig:system_model}
\end{figure}

As illustrated in Fig.~\ref{fig:system_model}, we consider a downlink
RSMA-enabled PASS, where an access point (AP) serves $K$ single-antenna
legitimate users in the presence of a single-antenna eavesdropper (Eve).
The AP feeds a dielectric waveguide through a single feed point, with
$N$ PAs deployed along the waveguide. Let
$\mathcal{K}\triangleq\{1,\ldots,K\}$ and
$\mathcal{N}\triangleq\{1,\ldots,N\}$ denote the sets of legitimate
users and PAs, respectively.

\subsection{Network Architecture}
\label{subsec:network_architecture}

A three-dimensional Cartesian coordinate system is adopted to
characterize the network geometry. Legitimate user $k\in\mathcal{K}$
and Eve are located at
$\mathbf{u}_k=(u_{k,x},u_{k,y},0)$ and
$\mathbf{u}_e=(u_{e,x},u_{e,y},0)$, respectively, within a rectangular
$L_x\times L_y$ service region. Their locations are assumed to be known
at the AP during the considered transmission interval. The dielectric
waveguide, of length $L_x$, is installed parallel to the $x$-axis at
height $H$, with the feed point located at
$\mathbf{p}_0=(0,0,H)$. PA $n\in\mathcal{N}$ is located at
$\mathbf{p}_n=(x_n,0,H)$, where $x_n$ denotes its adjustable position
along the waveguide. The PAs are indexed according to their positions,
such that
$0\leq x_1<\cdots<x_N\leq L_x$, with adjacent PAs satisfying
$x_n-x_{n-1}\geq D_{\min}$, $n=2,\ldots,N$, where $D_{\min}>0$
denotes the minimum allowable inter-PA spacing. The PA positions are
collected in $\mathbf{x}\triangleq[x_1,\ldots,x_N]^T$.

\subsection{PASS-Aided Downlink Channel Model}
\label{subsec:pass_channel_model}

We consider a line-of-sight (LoS)-dominant propagation model between
the PAs and the receivers \cite{11595855, 11481959}. For receiver
$q\in\mathcal{K}\cup\{e\}$, the distance from PA $n$ is
\begin{equation}
d_{q,n}(x_n)
\triangleq
\|\mathbf{u}_q-\mathbf{p}_n\|_2
=
\sqrt{(u_{q,x}-x_n)^2+u_{q,y}^2+H^2}.
\label{eq:pa_receiver_distance}
\end{equation}
Accordingly, the free-space channel coefficient between PA $n$ and
receiver $q$ is modeled as
\begin{equation}
h_{q,n}(x_n)
=
\frac{\lambda}{4\pi d_{q,n}(x_n)}
\exp\!\left(
-j\frac{2\pi}{\lambda}d_{q,n}(x_n)
\right),
\label{eq:free_space_channel}
\end{equation}
where $\lambda$ denotes the free-space carrier wavelength.
The guided-wave response from the feed point to PA $n$, accounting
for attenuation and phase accumulation, is modeled as
\begin{equation}
\phi_n(x_n)
=
\frac{1}{\sqrt{N}}
\exp(-\alpha_{\rm g}x_n)
\exp\!\left(
-j\frac{2\pi}{\lambda_{\rm g}}x_n
\right),
\label{eq:guided_wave_response}
\end{equation}
where $\alpha_{\rm g}\geq0$ denotes the in-waveguide amplitude
attenuation coefficient and $\lambda_{\rm g}$ denotes the guided
wavelength. The factor
$1/\sqrt{N}$ normalizes the aggregate PA response with respect to $N$.
The equivalent channel from the PASS feed point to receiver $q$ is
then given by the coherent superposition of the PA contributions as
\begin{equation}
g_q(\mathbf{x})
=
\sum_{n\in\mathcal{N}}
h_{q,n}(x_n)\phi_n(x_n),
\qquad q\in\mathcal{K}\cup\{e\}.
\label{eq:equivalent_channel}
\end{equation}
For notational simplicity, the dependence of $g_q(\mathbf{x})$ on
$\mathbf{x}$ is omitted whenever no ambiguity arises.

\subsection{Downlink RSMA Signal Model}
\label{subsec:rsma_signal_model}

Following the one-layer RSMA framework, the message intended for each
legitimate user is split into common and private parts. The common
parts are jointly encoded into a common stream $s_c$, while the private
parts are independently encoded into $s_k$, $k\in\mathcal{K}$. All
streams are mutually independent, zero-mean, and unit-power, i.e.,
$\mathbb{E}[|s_c|^2]=\mathbb{E}[|s_k|^2]=1$.

The composite baseband signal injected into the waveguide is
\begin{equation}
s
=
\sqrt{P_c}s_c
+
\sum_{k\in\mathcal{K}}\sqrt{P_k}s_k,
\label{eq:rsma_signal}
\end{equation}
where $P_c\geq0$ and $P_k\geq0$ denote the powers allocated to the
common stream and private stream $k$, respectively, subject to
\begin{equation}
P_c+\sum_{k\in\mathcal{K}}P_k\leq P_{\max},
\label{eq:total_power_constraint}
\end{equation}
where $P_{\max}$ denotes the maximum transmit-power budget.

Using the equivalent PASS channel in
\eqref{eq:equivalent_channel}, the received signal at receiver
$q\in\mathcal{K}\cup\{e\}$ is
\begin{equation}
y_q=g_q(\mathbf{x})s+n_q,
\label{eq:received_signal}
\end{equation}
where $n_q\sim\mathcal{CN}(0,\sigma_q^2)$ denotes additive white
Gaussian noise.

\subsection{Achievable and Secrecy Rates}
\label{subsec:secrecy_rates}

Each legitimate user first decodes the common stream while treating all
private streams as interference. Accordingly, the common-stream SINR
at user $k$ is
\begin{equation}
\gamma_{c,k}
=
\frac{|g_k|^2P_c}
{|g_k|^2\sum_{j\in\mathcal{K}}P_j+\sigma_k^2},
\label{eq:common_sinr}
\end{equation}
with achievable rate
$R_{c,k}\triangleq\log_2(1+\gamma_{c,k})$. Since the common stream
must be decoded by all legitimate users, its achievable rate is
\begin{equation}
R_c=\min_{k\in\mathcal{K}}R_{c,k}.
\label{eq:common_rate}
\end{equation}
After successfully decoding and canceling the common stream, user $k$
decodes its private stream with SINR
\begin{equation}
\gamma_k
=
\frac{|g_k|^2P_k}
{|g_k|^2\sum_{\substack{j\in\mathcal{K}\\j\neq k}}P_j+\sigma_k^2},
\label{eq:private_sinr}
\end{equation}
and achievable rate
$R_k\triangleq\log_2(1+\gamma_k)$.

Eve may also attempt to decode the common stream. Its corresponding
SINR and achievable rate are
\begin{equation}
\gamma_{e,c}
=
\frac{|g_e|^2P_c}
{|g_e|^2\sum_{j\in\mathcal{K}}P_j+\sigma_e^2},
\label{eq:eve_common_sinr}
\end{equation}
with $R_{e,c}\triangleq\log_2(1+\gamma_{e,c})$.
To prevent Eve from canceling the common stream prior to private-stream
decoding, we impose a positive decoding margin $\epsilon_s>0$ such that
\begin{equation}
R_{e,c}+\epsilon_s\leq R_c.
\label{eq:eve_common_constraint}
\end{equation}
Hence, the common stream remains as interference when Eve attempts to
decode private stream $k$, yielding
\begin{equation}
\gamma_{e,k}
=
\frac{|g_e|^2P_k}
{|g_e|^2
\left(
P_c+\sum_{\substack{j\in\mathcal{K}\\j\neq k}}P_j
\right)
+\sigma_e^2},
\label{eq:eve_private_sinr}
\end{equation}
with
$R_{e,k}\triangleq\log_2(1+\gamma_{e,k})$.
The corresponding private-stream secrecy rate is
\begin{equation}
R_{s,k}^{p}
=
[R_k-R_{e,k}]^+,
\qquad k\in\mathcal{K},
\label{eq:private_secrecy_rate}
\end{equation}
where $[z]^+\triangleq\max\{z,0\}$.

Under \eqref{eq:eve_common_constraint}, the achievable secrecy rate of
the common stream is
\begin{equation}
R_{s,c}
=
R_c-R_{e,c}.
\label{eq:common_secrecy_rate}
\end{equation}
Let $C_k\geq0$ denote the portion of this secure common rate allocated
to user $k$. The common-rate allocation therefore satisfies
$
\sum_{k\in\mathcal{K}}C_k\leq R_{s,c}.
\label{eq:common_rate_allocation}
$
The total secrecy rate of user $k$ is then
$
R_{s,k}=C_k+R_{s,k}^{p},
\label{eq:user_secrecy_rate}
$
and the resulting SSR is
\begin{equation}
R_{\rm sec}
=
\sum_{k\in\mathcal{K}}
\left(C_k+R_{s,k}^{p}\right).
\label{eq:ssr}
\end{equation}

\subsection{Problem Formulation}
\label{subsec:problem_formulation}

We jointly optimize the RSMA power allocation, secure common-rate
allocation, and PA positions to maximize the SSR. Let
$\mathbf{P}\triangleq[P_c,P_1,\ldots,P_K]^T$ and
$\mathbf{C}\triangleq[C_1,\ldots,C_K]^T$. The resulting problem is
formulated as
\begin{subequations}
\label{prob:P0}
\begin{align}
\text{(P0)}\quad
\max_{\mathbf{P},\mathbf{C},\mathbf{x}}\quad
& \sum_{k\in\mathcal{K}}\left(C_k+R_{s,k}^{p}\right)
\label{prob:P0_obj}\\
\mathrm{s.t.}\quad
& C_k+R_{s,k}^{p}\geq R_{s,k}^{\min},\quad \forall k\in\mathcal{K},
\label{prob:P0_qos}\\
& \sum_{j\in\mathcal{K}}C_j+R_{e,c}\leq R_{c,k},\quad \forall k\in\mathcal{K},
\label{prob:P0_common}\\
& R_{e,c}+\epsilon_s\leq R_{c,k},\quad \forall k\in\mathcal{K},
\label{prob:P0_eve}\\
& P_c+\sum_{k\in\mathcal{K}}P_k\leq P_{\max},
\label{prob:P0_power}\\
& P_c\geq0,\quad P_k\geq0,\quad C_k\geq0,\ \forall k\in\mathcal{K},
\label{prob:P0_nonnegative}\\
& 0\leq x_1<\cdots<x_N\leq L_x,
\label{prob:P0_position}\\
& x_n-x_{n-1}\geq D_{\min},\quad n=2,\ldots,N.
\label{prob:P0_spacing}
\end{align}
\end{subequations}
Here, $R_{s,k}^{\min}$ denotes the minimum secrecy-rate requirement of
user $k$. Constraint \eqref{prob:P0_qos} enforces the per-user secrecy
requirement, while \eqref{prob:P0_common} ensures that the allocated
secure common rate does not exceed the common-stream secrecy rate.
Constraint \eqref{prob:P0_eve} prevents Eve from canceling the common
stream before attempting private-stream decoding. Finally,
\eqref{prob:P0_power}--\eqref{prob:P0_spacing} impose the transmit-power,
nonnegativity, and PA-placement constraints.

Problem \textnormal{(P0)} is nonconvex due to the coupled interference
terms and the nonlinear dependence of the PASS channels on the PA
positions. We therefore develop an AO framework that alternately
optimizes the RSMA resource allocation $(\mathbf{P},\mathbf{C})$ and
PA positions $\mathbf{x}$.

\section{Proposed Optimization Framework}
\label{sec:optimization}

The two variable blocks in the AO framework are handled differently.
For fixed PA positions, the RSMA resource-allocation subproblem is
solved via SCA. For fixed RSMA
resource allocation, the PA positions are updated via PGA. The two steps are detailed next.

\subsection{RSMA Resource-Allocation Optimization}
\label{subsec:resource_sca}

For fixed $\mathbf{x}$, let
$a_k\triangleq|g_k(\mathbf{x})|^2$,
$a_e\triangleq|g_e(\mathbf{x})|^2$, and
$S\triangleq\sum_{j\in\mathcal{K}}P_j$. Before constructing the SCA subproblem, the following result simplifies
the private-stream secrecy rates.

\textit{Lemma 1:} Under constraint \eqref{prob:P0_eve},
$R_k\geq R_{e,k}$ for all $k\in\mathcal{K}$. Hence,
$R_{s,k}^{p}=R_k-R_{e,k}$ over the feasible set of
\textnormal{(P0)}.

\textit{Proof:} See Appendix~A.

Using Lemma~1, define the private-stream secrecy-rate difference as
$
\Delta_k(\mathbf{P})
\triangleq
R_k-R_{e,k}.
\label{eq:delta_definition}
$
For fixed $\mathbf{x}$, it can be expressed as
\begin{align}
\Delta_k(\mathbf{P})
={}&
\log_2(a_kS+\sigma_k^2)
+\log_2\!\bigl(a_e(P_c+S-P_k)+\sigma_e^2\bigr)
\nonumber\\
&-\log_2\!\bigl(a_k(S-P_k)+\sigma_k^2\bigr)
-\log_2\!\bigl(a_e(P_c+S)+\sigma_e^2\bigr).
\label{eq:private_secrecy_dc}
\end{align}
Although each logarithmic term has an affine argument in
$\mathbf{P}$, \eqref{eq:private_secrecy_dc} is a difference of concave
functions and is therefore generally nonconcave.

For $f(z)\triangleq\log_2 z$, concavity implies the global affine
upper bound
\begin{equation}
f(z)
\leq
f(z^{(r)})
+
\frac{z-z^{(r)}}{z^{(r)}\ln 2}
\triangleq
\widehat f(z;z^{(r)}),
\label{eq:log_upper_bound}
\end{equation}
where $z^{(r)}>0$ denotes the value at the current SCA iterate.
Applying \eqref{eq:log_upper_bound} to the two subtracted logarithmic
terms in \eqref{eq:private_secrecy_dc} yields the concave lower bound
\begin{align}
\underline{\Delta}_k^{(r)}(\mathbf{P})
={}&
\log_2(a_kS+\sigma_k^2)
+\log_2\!\bigl(a_e(P_c+S-P_k)+\sigma_e^2\bigr)
\nonumber\\
&-\widehat f\!\left(
a_k(S-P_k)+\sigma_k^2;\
a_k(S^{(r)}-P_k^{(r)})+\sigma_k^2
\right)
\nonumber\\
&-\widehat f\!\left(
a_e(P_c+S)+\sigma_e^2;\
a_e(P_c^{(r)}+S^{(r)})+\sigma_e^2
\right),
\label{eq:private_secrecy_lower_bound}
\end{align}
where
$S^{(r)}\triangleq\sum_{j\in\mathcal{K}}P_j^{(r)}$.
Hence,
$
\underline{\Delta}_k^{(r)}(\mathbf{P})
\leq
\Delta_k(\mathbf{P}),
\label{eq:delta_lower_bound_property}
$
with equality at $\mathbf{P}=\mathbf{P}^{(r)}$.
Next, define
$
G_k(\mathbf{P})
\triangleq
R_{e,c}-R_{c,k}.
\label{eq:G_definition}
$
Using the common-stream rates, $G_k(\mathbf{P})$ can be written as
\begin{align}
G_k(\mathbf{P})
={}&
\log_2\!\bigl(a_e(P_c+S)+\sigma_e^2\bigr)
-\log_2(a_eS+\sigma_e^2)
\nonumber\\
&-\log_2\!\bigl(a_k(P_c+S)+\sigma_k^2\bigr)
+\log_2(a_kS+\sigma_k^2).
\label{eq:G_dc}
\end{align}
By upper-bounding the two positively signed concave logarithmic terms
using \eqref{eq:log_upper_bound}, we obtain the convex upper bound
\begin{align}
\widehat G_k^{(r)}(\mathbf{P})
={}&
\widehat f\!\left(
a_e(P_c+S)+\sigma_e^2;\
a_e(P_c^{(r)}+S^{(r)})+\sigma_e^2
\right)
\nonumber\\
&-\log_2(a_eS+\sigma_e^2)
-\log_2\!\bigl(a_k(P_c+S)+\sigma_k^2\bigr)
\nonumber\\
&+\widehat f\!\left(
a_kS+\sigma_k^2;\
a_kS^{(r)}+\sigma_k^2
\right),
\label{eq:G_upper_bound}
\end{align}
which satisfies
$
G_k(\mathbf{P})
\leq
\widehat G_k^{(r)}(\mathbf{P}),
\label{eq:G_upper_bound_property}
$
with equality at $\mathbf{P}=\mathbf{P}^{(r)}$.

Introducing auxiliary variables
$\boldsymbol{\rho}\triangleq[\rho_1,\ldots,\rho_K]^T$ for the
private-stream secrecy rates, the resource-allocation variables at SCA
iteration $r$ are updated by solving
\begin{subequations}
\label{prob:P1}
\begin{align}
\text{(P1)}\quad
\max_{\mathbf{P},\mathbf{C},\boldsymbol{\rho}}\quad
& \sum_{k\in\mathcal{K}}(C_k+\rho_k)
\label{prob:P1_obj}\\
\mathrm{s.t.}\quad
& C_k+\rho_k\geq R_{s,k}^{\min},
\quad \forall k\in\mathcal{K},
\label{prob:P1_qos}\\
& 0\leq\rho_k\leq
\underline{\Delta}_k^{(r)}(\mathbf{P}),
\quad \forall k\in\mathcal{K},
\label{prob:P1_private}\\
& \sum_{j\in\mathcal{K}}C_j+
\widehat G_k^{(r)}(\mathbf{P})\leq0,
\quad \forall k\in\mathcal{K},
\label{prob:P1_common}\\
& \widehat G_k^{(r)}(\mathbf{P})+\epsilon_s\leq0,
\quad \forall k\in\mathcal{K},
\label{prob:P1_eve}\\
& P_c+\sum_{k\in\mathcal{K}}P_k\leq P_{\max},
\label{prob:P1_power}\\
& \mathbf{P}\succeq\mathbf{0},\qquad
\mathbf{C}\succeq\mathbf{0}.
\label{prob:P1_nonnegative}
\end{align}
\end{subequations}
Problem \textnormal{(P1)} is convex since
$\underline{\Delta}_k^{(r)}(\mathbf{P})$ is concave and
$\widehat G_k^{(r)}(\mathbf{P})$ is convex. Moreover, the adopted
bounds are tight at the current iterate and conservatively approximate
the corresponding constraints of \textnormal{(P0)}. Hence, starting
from a feasible point, solving \textnormal{(P1)} yields a feasible
resource-allocation update with a nondecreasing objective value.
The SCA iterations are repeated until convergence.

\subsection{PA-Position Optimization}
\label{subsec:position}

For fixed $(\mathbf{P},\mathbf{C})$, the PA positions are optimized
while keeping the RSMA resource allocation unchanged. Since
$\mathbf{x}$ affects both the propagation losses and the guided- and
free-space phases, the resulting subproblem remains highly nonconvex.
We therefore employ PGA with backtracking
line search.

Since $\sum_{k\in\mathcal{K}}C_k$ is constant in this subproblem, the
position-dependent part of the SSR is
\begin{equation}
F(\mathbf{x})
\triangleq
\sum_{k\in\mathcal{K}}
\bigl[R_k(\mathbf{x})-R_{e,k}(\mathbf{x})\bigr]^+ .
\label{eq:position_objective}
\end{equation}
By Lemma~1, the positive-part operator is inactive at every feasible
point of \textnormal{(P0)}. At position iteration $t$, the $n$th
component of the gradient is numerically approximated as
\begin{equation}
[\nabla F(\mathbf{x}^{(t)})]_n
\simeq
\frac{F(\mathbf{x}^{(t)}+\delta\mathbf{e}_n)
-F(\mathbf{x}^{(t)}-\delta\mathbf{e}_n)}
{2\delta},
\label{eq:fd_gradient}
\end{equation}
where $\delta>0$ is the finite-difference interval and $\mathbf{e}_n$
denotes the $n$th canonical basis vector. Centered differences are used
when both perturbations satisfy the PA-placement constraints; otherwise,
the corresponding one-sided difference is employed.

For a step size $\eta_t>0$, a candidate position vector is obtained as
\begin{equation}
\widetilde{\mathbf{x}}^{(t+1)}
=
\Pi_{\mathcal{X}}
\!\left(
\mathbf{x}^{(t)}
+\eta_t\nabla F(\mathbf{x}^{(t)})
\right),
\label{eq:pga_update}
\end{equation}
where $\Pi_{\mathcal{X}}(\cdot)$ denotes the Euclidean projection onto
the feasible PA-placement region, i.e.,
\begin{equation}
\begin{aligned}
\Pi_{\mathcal{X}}(\mathbf{z})
\triangleq\;&
\arg\min_{\mathbf{x}}\ \|\mathbf{x}-\mathbf{z}\|_2^2\\
\mathrm{s.t.}\;&
0\leq x_1,\quad x_N\leq L_x,\\
&x_n-x_{n-1}\geq D_{\min},\quad n=2,\ldots,N .
\end{aligned}
\label{eq:position_projection}
\end{equation}
The projection in \eqref{eq:position_projection} is a convex quadratic
program with linear constraints. Starting from a feasible position
vector, the step size $\eta_t$ is reduced via backtracking until
$\widetilde{\mathbf{x}}^{(t+1)}$ satisfies all position-dependent
constraints of \textnormal{(P0)} and
$
F(\widetilde{\mathbf{x}}^{(t+1)})
\geq
F(\mathbf{x}^{(t)}).
\label{eq:pga_acceptance}
$
If no acceptable step is found, the current position vector is retained.

The overall procedure is summarized in Algorithm~\ref{alg:AO}.

\begin{algorithm}[t]
\caption{AO Algorithm for Secure RSMA-Enabled PASS}
\label{alg:AO}
\begin{algorithmic}
\State Initialize a feasible
$(\mathbf{P}^{(0)},\mathbf{C}^{(0)},\mathbf{x}^{(0)})$; set $i=0$.
\Repeat
    \State Fix $\mathbf{x}^{(i)}$; set
    $(\mathbf{P}_{\rm s}^{(0)},\mathbf{C}_{\rm s}^{(0)})
    =(\mathbf{P}^{(i)},\mathbf{C}^{(i)})$ and $r=0$.
    \Repeat
        \State Construct $\underline{\Delta}_k^{(r)}$ and
        $\widehat{G}_k^{(r)}$, $\forall k\in\mathcal{K}$.
        \State Solve \textnormal{(P1)} to obtain
        $(\mathbf{P}_{\rm s}^{(r+1)},\mathbf{C}_{\rm s}^{(r+1)})$;
        set $r\gets r+1$.
    \Until{SCA convergence}
    \State Set
    $(\mathbf{P}^{(i+1)},\mathbf{C}^{(i+1)})
    =(\mathbf{P}_{\rm s}^{(r)},\mathbf{C}_{\rm s}^{(r)})$.
    \State Set $\mathbf{x}_{\rm p}^{(0)}=\mathbf{x}^{(i)}$ and $t=0$.
    \Repeat
        \State Evaluate $\nabla F(\mathbf{x}_{\rm p}^{(t)})$
        via \eqref{eq:fd_gradient}.
        \State Update $\mathbf{x}_{\rm p}^{(t+1)}$ via
        \eqref{eq:pga_update} with backtracking; $t\gets t+1$.
    \Until{PGA convergence}
    \State Set $\mathbf{x}^{(i+1)}=\mathbf{x}_{\rm p}^{(t)}$;
    set $i\gets i+1$.
\Until{AO convergence}
\State \textbf{Output:}
$(\mathbf{P}^{\star},\mathbf{C}^{\star},\mathbf{x}^{\star})$.
\end{algorithmic}
\end{algorithm}

The SCA step yields a nondecreasing SSR since its surrogates are tight
at the current iterate, while the backtracking procedure accepts only
feasible PGA updates that do not decrease the SSR. Hence,
Algorithm~\ref{alg:AO} generates a nondecreasing SSR sequence, which
converges since the SSR is upper bounded under the finite transmit-power
budget.

\subsection{Computational Complexity}
\label{subsec:complexity}

Let $I_{\rm AO}$, $I_{\rm SCA}$, and $I_{\rm PGA}$ denote the numbers
of AO, SCA, and PGA iterations, respectively. Problem
\textnormal{(P1)} involves $3K+1$ optimization variables; let
$\mathcal{C}_{\rm P1}(K)$ denote the complexity of solving it.
For the position update, evaluating the finite-difference gradient
requires $2N$ objective evaluations, each involving $K+1$ effective
channels with $N$ PA contributions, resulting in
$\mathcal{O}(KN^2)$ complexity. Denoting the complexity of the
projection in \eqref{eq:position_projection} by
$\mathcal{C}_{\rm proj}(N)$, the overall complexity is
$
\mathcal{O}\!\left(
I_{\rm AO}
\left[
I_{\rm SCA}\mathcal{C}_{\rm P1}(K)
+
I_{\rm PGA}
\bigl(
KN^2+\mathcal{C}_{\rm proj}(N)
\bigr)
\right]
\right).
\label{eq:overall_complexity}
$

\section{Numerical Results}
\label{sec:numerical_results}

In this section, numerical results are presented to evaluate the
secrecy performance of the proposed RSMA-enabled PASS design. The simulation setup follows the related literature
\cite{11595855, 11481959,Yang2026NOMAPASS}. Unless
otherwise specified, we consider $K=3$ legitimate users and $N=8$ PAs
deployed over a waveguide of length $L_x=20$ m at height $H=3$ m,
with service-region width $L_y=10$ m. The carrier frequency is
$f_c=28$ GHz, the guided wavelength is $\lambda_g=8$ mm, and the
waveguide amplitude attenuation coefficient is
$\alpha_g=0.01$ m$^{-1}$. The noise power at all receivers is
$-90$ dBm, and $D_{\min}=\lambda/2$.
Furthermore, $P_{\max}=10$ dBm,
$R_{s,k}^{\min}=0.1$ bit/s/Hz, and $\epsilon_s=0.01$ bit/s/Hz.
The reported results are averaged over
$400$ independent location realizations, where the legitimate users
and Eve are independently and uniformly distributed over the service
region and remain fixed during each optimization run.

For comparison, the following benchmark schemes are considered:
\begin{itemize}
    \item \textbf{Fixed-Position RSMA-PASS:} The PAs are uniformly
    spaced and fixed, while the RSMA power and secure common-rate
    allocations are optimized.

    \item \textbf{Equal-Power RSMA-PASS:} The common and private
    streams are assigned equal power, i.e.,
    $P_c=P_k=P_{\max}/(K+1)$, $\forall k\in\mathcal{K}$, while the
    secure common-rate allocation and PA positions are optimized.

    \item \textbf{Private-Only PASS:} The common stream is disabled
    $(P_c=0,\mathbf{C}=\mathbf{0})$, while the private-stream powers
    and PA positions are jointly optimized; the common-stream
    constraints are omitted.
\end{itemize}

\begin{figure*}[t]
    \centering
    \subfloat[]
    {\includegraphics[
    width=0.242\textwidth,
    height=2.8cm]{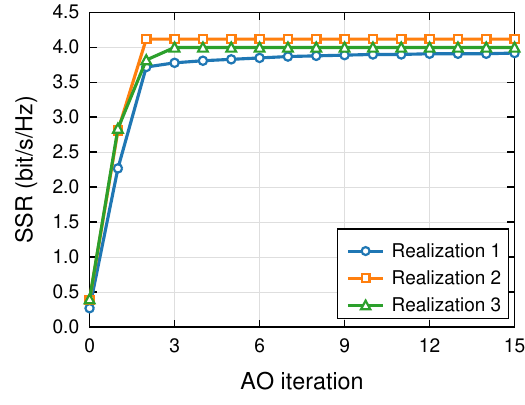}
    \label{fig:convergence}}
    \hfill
    \subfloat[]
    {\includegraphics[
    width=0.242\textwidth,
    height=2.8cm]{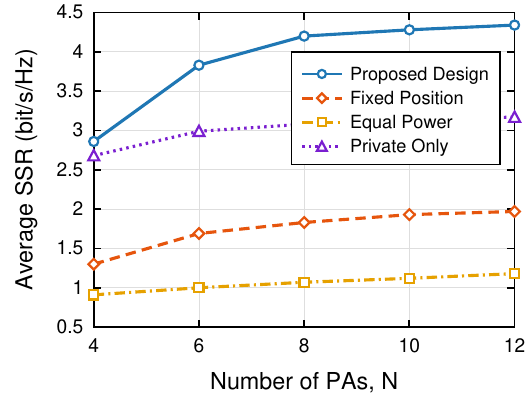}
    \label{fig:ssr_vs_N}}
    \hfill
    \subfloat[]
    {\includegraphics[
    width=0.242\textwidth,
    height=2.8cm]{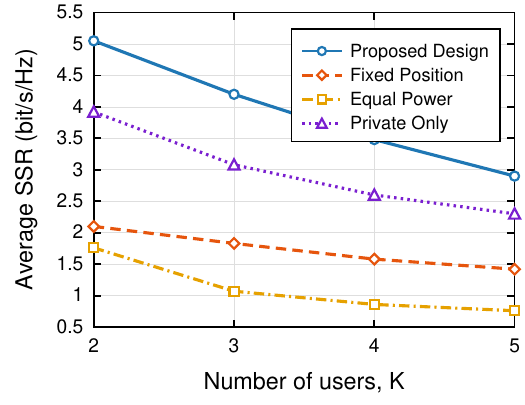}
    \label{fig:ssr_vs_K}}
    \hfill
    \subfloat[]
    {\includegraphics[
    width=0.242\textwidth,
    height=2.8cm
]{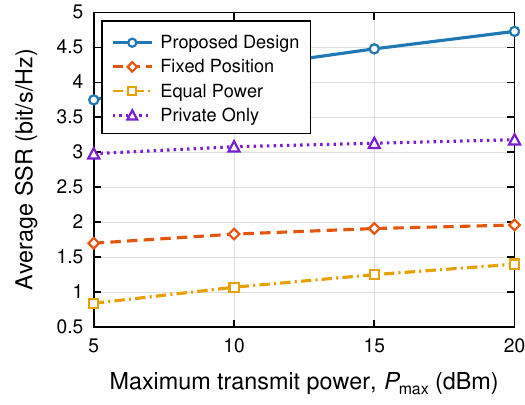}
    \label{fig:ssr_vs_power}}
    \caption{Performance evaluation of the proposed design:
    (a) convergence behavior; (b) average SSR versus the number of PAs;
    (c) average SSR versus the number of legitimate users; and
    (d) average SSR versus the maximum transmit power.}
    \label{fig:numerical_results}
\end{figure*}

Fig.~\ref{fig:convergence} shows the convergence behavior of the
proposed AO algorithm for three representative location realizations.
For all realizations, the SSR increases monotonically and stabilizes
within a few AO iterations, demonstrating the fast convergence of the
proposed algorithm.

Fig.~\ref{fig:ssr_vs_N} shows the average SSR versus the number of PAs.
The proposed design consistently outperforms all benchmark schemes,
with the SSR generally increasing as more PAs are deployed. The
performance gaps over the fixed-position and equal-power schemes
highlight the importance of jointly optimizing the PA positions and
RSMA resource allocation, while the gain over the private-only scheme
demonstrates the benefit of RSMA for secure PASS transmission. The
diminishing improvement at larger $N$ indicates that the additional
spatial gains gradually saturate.

Fig.~\ref{fig:ssr_vs_K} shows the average SSR versus the number of
legitimate users. As $K$ increases, the SSR decreases for all schemes,
as the available transmit power must be shared among more users while
multiuser interference and secrecy requirements become more
restrictive. Nevertheless, the proposed design consistently achieves
the highest SSR. The gain over the private-only scheme further
demonstrates the effectiveness of RSMA for secure multiuser PASS
transmission.

Fig.~\ref{fig:ssr_vs_power} shows the average SSR versus the maximum
transmit power. As $P_{\max}$ increases, the SSR improves for all
considered schemes due to the increased available transmit power.
The proposed design consistently achieves the highest SSR, with a more
pronounced performance advantage at higher power budgets,
demonstrating the effectiveness of joint RSMA resource and PA-position
optimization in exploiting the available transmit power.

\section{Conclusion}
\label{sec:conclusion}

This paper investigated SSR maximization for a downlink RSMA-enabled
single-waveguide PASS by jointly optimizing the RSMA power allocation,
secure common-rate allocation, and PA positions. An AO framework
combining SCA-based resource allocation and feasibility-preserving PGA
was developed to address the resulting nonconvex problem. Numerical
results showed consistent SSR gains over the fixed-position,
equal-power, and private-only benchmarks, demonstrating the benefits of
joint RSMA resource and PA-position optimization for secure PASS
transmission.

\appendices
\section{}
\label{app:lemma1}

Since $\epsilon_s>0$, \eqref{prob:P0_eve} implies
$R_{c,k}>R_{e,c}$. Moreover, any feasible point must satisfy $P_c>0$;
otherwise, $R_{c,k}=R_{e,c}=0$, contradicting
\eqref{prob:P0_eve}. Hence,
\begin{equation}
\frac{a_kP_c}{a_kS+\sigma_k^2}
>
\frac{a_eP_c}{a_eS+\sigma_e^2}
\ \Longrightarrow\
a_k\sigma_e^2>a_e\sigma_k^2.
\label{eq:channel_advantage}
\end{equation}
For $P_k>0$,
\begin{align}
&a_k[a_e(P_c+S-P_k)+\sigma_e^2]
-a_e[a_k(S-P_k)+\sigma_k^2]
\nonumber\\
&=a_ka_eP_c+a_k\sigma_e^2-a_e\sigma_k^2>0,
\end{align}
where the inequality follows from \eqref{eq:channel_advantage}.
Thus, $\gamma_k>\gamma_{e,k}$ and consequently $R_k>R_{e,k}$.
For $P_k=0$, $R_k=R_{e,k}=0$. Therefore,
$R_k\geq R_{e,k}$ for all $k\in\mathcal{K}$, completing the proof.

	\bibliographystyle{IEEEtran}
	\bibliography{references}

@ARTICLE{11222687,
  author={Yang, Zheng and others},
  journal={IEEE Wireless Commun.}, 
  title={Pinching Antennas: Principles, Applications and Challenges}, 
  year={2026},
  month={Apr.},
  volume={33},
  number={2},
  pages={175-184}}

@ARTICLE{11625963,
  author={Zeng, Ming and others},
  journal={IEEE Wireless Commun.}, 
  title={Resource Allocation for Pinching-Antenna Systems: State-of-the-Art, Key Techniques and Open Issues}, 
  year={2026},
  month={Jul.},
  volume={},
  number={},
  pages={1-7}}

@ARTICLE{11029492,
  author={Zhou, Ziwu and others},
  journal={IEEE Wireless Commun. Lett.}, 
  title={Sum-Rate Maximization for {NOMA}-Assisted Pinching-Antenna Systems}, 
  year={2025},
  month={Sep.},
  volume={14},
  number={9},
  pages={2728-2732}}

@ARTICLE{11568577,
  author={Xue, Songtao and others},
  journal={IEEE Trans. Commun.}, 
  title={Resource Allocation for Pinching-Antenna Systems ({PASS})-Enabled {NOMA} Communications}, 
  year={2026},
  month={Jun.},
  volume={74},
  number={},
  pages={10246-10260}}

@ARTICLE{11202497,
  author={Zhu, Guangyu and others},
  journal={IEEE Trans. Commun.}, 
  title={Pinching-Antenna Systems ({PASS})-Enabled Secure Wireless Communications}, 
  year={2025},
  month={Oct.},
  volume={74},
  number={},
  pages={490-505}}

@ARTICLE{11558589,
  author={Cui, Huanxi and others},
  journal={IEEE Trans. Wireless Commun.}, 
  title={Pinching Antenna and Movable Signal Enabled Wireless Communication}, 
  year={2026},
  month={Jun.},
  volume={25},
  number={},
  pages={18393-18409}}

@ARTICLE{11597999,
  author={Cui, Huanxi and others},
  journal={IEEE Trans. Wireless Commun.}, 
  title={Movable-Signal and Pinching-Antenna for Integrated Sensing and Communications}, 
  year={2026},
  month={Jul.},
  volume={25},
  number={},
  pages={19325-19341}}

@article{Pakravan2026PASSRSMA,
  author  = {Pakravan, Saeid and others},
  title   = {{PASS}-Assisted {RSMA} under Imperfect {SIC}: Joint Antenna Activation and Resource Allocation},
  journal = {arXiv preprint arXiv:2606.28921},
  year    = {2026},
  month={Jun.},
}

@inproceedings{ref_conference,
  author    = {Pakravan, Saeid and others},
  title     = {Pinching Antenna System-Assisted Hybrid {AirComp}-{NOMA} Uplink: Joint Precoding and Antenna Placement Optimization},
  booktitle = {Proc. IEEE Int. Conf. Wireless Mobile Comput., Netw. Commun. (WiMob)},
pages={},
  year={},
  month = {},
  note={{A}vignon, {F}rance, pp. 1--6, Oct. 2026.}
}

@ARTICLE{11481959,
  author={Pakravan, Saeid and others},
  journal={IEEE Trans. Veh. Technol.}, 
  title={{AI}-Empowered Resource Allocation for Wirelessly Powered Pinching-Antenna Systems}, 
  year={2026},
  month={Apr.},
  volume={75},
  number={9},
  pages={21820-21825}}

@ARTICLE{11595855,
  author={Feng, Hao and others},
  journal={IEEE Wireless Commun. Lett.}, 
  title={Hybrid Wireless-Fed Pinching-Antenna Systems With Residual Self-Interference-Aware Optimization}, 
  year={2026},
  month={Jul.},
  volume={15},
  number={},
  pages={4025-4029}}

@ARTICLE{9963667,
  author={Gao, Ying and others},
  journal={IEEE Commun. Lett.}, 
  title={Rate-Splitting Multiple Access for Intelligent Reflecting Surface-Aided Secure Transmission}, 
  year={2023},
  month={Feb.},
  volume={27},
  number={2},
  pages={482-486}}

@ARTICLE{11205176,
  author={Zhong, Yuan and others},
  journal={IEEE Wireless Commun. Lett.}, 
  title={Physical Layer Security for Pinching-Antenna Systems via Index and Directional Modulation}, 
  year={2025},
  month={Oct.},
  volume={15},
  number={},
  pages={230-234}}

@article{Yang2026NOMAPASS,
  author  = {Yang, Hui and others},
  title   = {{NOMA}-Assisted Multi-User Hybrid Wireless-Fed Pinching-Antenna Systems},
  journal = {arXiv preprint arXiv:2609.23698},
  year    = {2026},
  month={Sep.},
}

	\vfill
	
\end{document}